\documentclass[letterpaper,aps,reprint,prd,nofootinbib,noshowkeys,longbibliography,floatfix,superscriptaddress,preprintnumbers]{revtex4-1}
\usepackage{hyperref} 
\usepackage{amsmath}
\usepackage{amsfonts}
\usepackage{amsthm}
\usepackage{amssymb}
\usepackage{cancel}
\usepackage{graphicx}
\usepackage{feynmf}
\usepackage{bbding}
\usepackage{slashed}
\usepackage[sort&compress]{natbib}

\newcommand*\Temp{\tilde{T}} 
\newcommand*\Action{\tilde{S}} 
\newcommand{\mr}[1]{\mathrm{#1}}

\newcommand{\mc}[1]{\ensuremath{\mathcal #1}}

\begin{document}

\title{Worldline sphaleron for thermal Schwinger pair production}
\date{August 24, 2018}
\author{Oliver Gould}
\email{o.gould13@imperial.ac.uk}
\affiliation{Department of Physics, Imperial College London, SW7 2AZ, UK}
\affiliation{Helsinki Institute of Physics, University of Helsinki, FI-00014, Finland}
\preprint{IMPERIAL-TP-2018-OG-1}
\preprint{HIP-2018-17-TH}
\author{Arttu Rajantie}
\email{a.rajantie@imperial.ac.uk}
\affiliation{Department of Physics, Imperial College London, SW7 2AZ, UK}
\author{Cheng Xie}
\email{cxie@live.com}
\affiliation{Department of Physics, Imperial College London, SW7 2AZ, UK}


\begin{abstract}
With increasing temperatures, Schwinger pair production changes from a quantum tunnelling to a classical, thermal process, determined by a worldline sphaleron. We show this and calculate the corresponding rate of pair production for both spinor and scalar quantum electrodynamics, including the semiclassical prefactor. For electron-positron pair production from a thermal bath of photons and in the presence of an electric field, the rate we derive is faster than both perturbative photon fusion and the zero temperature Schwinger process. We work to all-orders in the coupling and hence our results are also relevant to the pair production of (strongly coupled) magnetic monopoles in heavy-ion collisions.
\end{abstract}

\maketitle

\section{Introduction}

In non-Abelian gauge theories, sphaleron processes, or thermal over-barrier transitions, have long been understood to dominate over quantum tunnelling transitions at high enough temperatures \cite{manton1983topology,klinkhamer1984saddle,kuzmin1985anomalous,arnold1987sphalerons}\footnote{The word \emph{sphaleron} was originally coined in Refs. \cite{manton1983topology,klinkhamer1984saddle}. It denotes a static, localised, unstable field configuration, which is a minimum of the energy functional in all directions in function space except one, where it is a maximum.}. The same is true, for example, in gravitational theories \cite{hawking1981supercooled}. On the other hand, sphalerons have been conspicuously absent from the study of Abelian gauge theories. In this paper, we make partial amends for this absence by finding a sphaleron in quantum electrodynamics (QED). We note that, unlike the corresponding quantum tunnelling transition at zero temperature, this sphaleron is not visible at any finite order in the loop expansion, which may explain why it has been missed in the past.

In the presence of an electric field, empty space is nonperturbatively unstable to decay into electron-positron pairs, called Schwinger pair production \cite{schwinger1951gauge}. At zero temperature, Schwinger pair production is determined by a circular worldline instanton \cite{affleck1981pair}. At nonzero temperatures the rate of this process is enhanced by the energy of the thermal bath, and the worldline instanton is deformed away from circular \cite{selivanov1986tunneling,ivlev1987tunneling,monin2010photon,brown2015schwinger,gould2017thermal}. At sufficiently high temperatures the process becomes essentially thermal and is determined by a worldline sphaleron (which, in Ref. \cite{gould2017thermal}, we referred to as the S instanton). In this paper we briefly review the derivation of the worldline sphaleron and then calculate the sphaleron rate, including the fluctuation prefactor.

At zero temperature, Schwinger pair production is a quantum tunnelling process visible at one loop. If we denote symbolically the interaction between the dynamical (as opposed to external) photon field and the charged particles as $J\cdot A$, then the loop expansion of the Schwinger rate, $\Gamma(E,T)$, takes the form,
\begin{equation}
 \Gamma(E,T) = \sum_{n=0}^{\infty} c_n\langle (J\cdot A)^n \rangle, \label{eq:loop_expansion}
\end{equation}
where $E$ is the magnitude of the electric field and $T$ is the temperature. The leading term, $c_0$, gives the one loop result, that of Schwinger \cite{schwinger1951gauge}, at zero temperature. There has been some disagreement in the literature about the thermal corrections to $c_0$. According to the worldline instanton calculation, at low temperatures there are no thermal corrections at one loop, so agreeing with Refs. \cite{cox1984finite,elmfors1994electromagnetic,gies1999qed} using alternative approaches. However at high enough temperatures, such that $T>gE/(2m)$,\footnote{Except where noted, we use the natural units common in high energy physics, where $c=\hbar=k_B=\epsilon_0=1$.} where $g$ is the charge coupling and $m$ is the mass of the (lightest) charged particles, the worldline instanton calculation suggests that the one loop calculation is no longer a consistent truncation of the problem and thermal corrections are expected at leading order \cite{selivanov1986tunneling,brown2015schwinger,medina2015schwinger,gould2017thermal}. See Refs. \cite{loewe1992thermal,ganguly1995thermal,kim2010nonperturbative} for alternative conclusions regarding the thermal, one loop correction. The next order term, $c_1\langle  J\cdot A \rangle$, has also been calculated, both at zero \cite{ritus1975lagrange,ritus1977connection,lebedev1984virial,ritus1998effective} and at nonzero but low temperature \cite{gies2000qed,monin2010photon}. Now, the presence of the worldline sphaleron requires the dynamical photon interaction, $J\cdot A$, in the exponent of the semiclassical expansion. Hence, it requires all orders in the loop expansion.

\begin{figure}
 \centering
  \includegraphics[width=1.0\columnwidth]{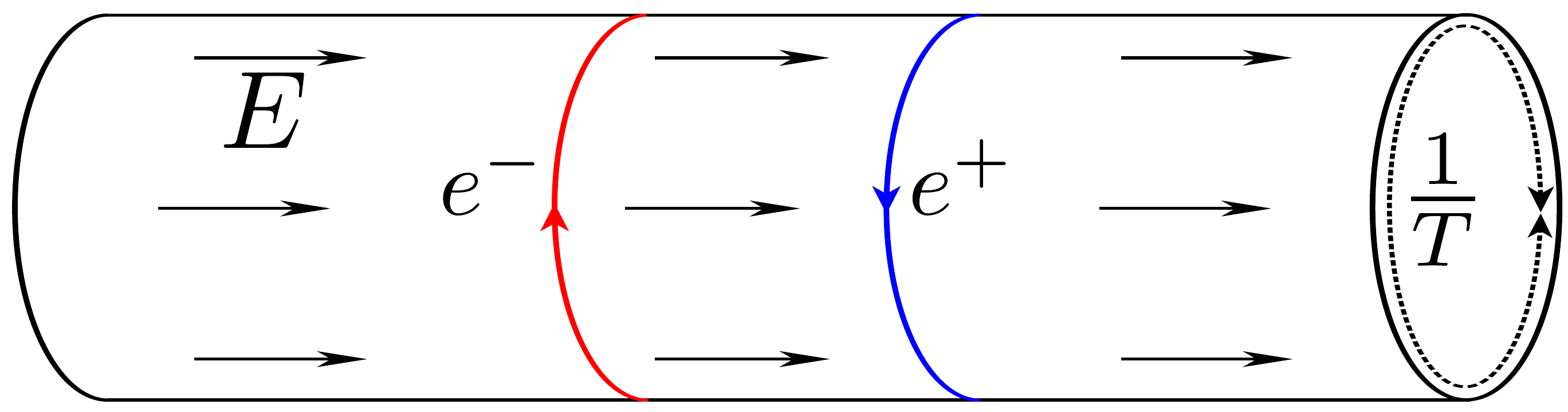}
  \caption[Worldline sphaleron.]{The simple worldline sphaleron, for the thermal Schwinger process. The Coulomb attraction between electron, $e^-$, and positron, $e^+$, is balanced by the external electric field, $E$, which pushes them apart. Pictured here in the Euclidean thermal approach, where the solution lives is $R^3\times S^1$, the circumference of the circle being the inverse temperature, $1/T$.}
  \label{fig:worldlineSphaleron}
\end{figure}

Despite this, the worldline sphaleron itself is rather simple. The electric version consists of an electron and a positron a finite distance apart, such that the attractive Coulomb force between them is balanced by the force of the external electric field pushing them apart (see Fig. \ref{fig:worldlineSphaleron}).

Following the work in Ref. \cite{gould2017thermal}, our calculations in this paper are carried out in both QED and scalar QED (SQED), making no assumptions as regards the magnitude of the charge coupling, $g$. We let $g$ range from infinitesimal to infinite. In SQED we will assume however that the scalar self-coupling is weak, $\lambda\ll 1$\footnote{Of course photon loops will generate this term. However, the term is a pointlike interaction between scalar loops (given no external legs) and, in the dilute instanton approximation that we will make, such loops are subdominant and are neglected.\label{small_lambda}}. We consider circumstances when the calculation is semiclassical and hence the rate is slow. In this regime the results are largely independent of many properties of the charged particles.

We should add a quick note on what we mean by our initial thermal state. We consider temperatures, $T$, much less than the mass of any charged particles, $m$. In this case there are no charged particles in the initial state, it is a thermal bath of photons. Upon turning on the external field, charged particles are produced by the thermal Schwinger process. After an initial transitory period, which depends on how quickly the electric field is turned on, and before the back-reaction of the charged particles becomes significant, there is a period during which charged particles are produced at a constant rate \cite{gavrilov2008one,hebenstreit2013simulating}. This rate is what we calculate. The process is relevant only for the lightest charged particle (either the electron or the lightest magnetic monopole), because the lightest particles will be produced exponentially more quickly than heavier particles and, once produced, their presence will cause Debye screening.

The chief result of this paper is the following, the sphaleron rate for thermal Schwinger pair production,
\begin{widetext}
\begin{equation}
 \Gamma(E,T)\approx \frac{(2s+1)^2T_{WS}\left(mT\right)^{3/2}}{(4\pi)^{3/2}\sin \left(\frac{\pi T_{WS} }{T }\right)\sinh^2\left(\frac{\pi T_{WS}}{\sqrt{2} T }\right)} \mr{e}^{-\frac{2m}{T}+\frac{\sqrt{g^3E/\pi}}{T}}
 \left[1+O\left(\frac{g E}{m^2},\frac{g^3 E}{m^2},\frac{T}{m},\frac{g^2T}{m},\frac{g T^2}{E}\right)\right],\label{eq:rate_intro}
\end{equation}
\end{widetext}
where $s$ is the spin of the charged particles (either zero or 1/2) and the expression applies for temperatures satisfying
\begin{equation}
 T>T_{WS}\approx\left(\frac{4 g E^3}{\pi^3 m^2}\right)^{1/4}. \label{eq:tws}
\end{equation}
Eqs. \eqref{eq:rate_intro} and \eqref{eq:tws} are valid to all orders in the coupling $g$, though only to leading order in the five different dimensionless combinations of parameters within the big $O$ symbol in Eq. \eqref{eq:rate_intro}. Hence all of these parameters must be small for our approximations to be good. This is naturally satisfied, for all but the last of these parameters, if the charged particles are sufficiently heavy.

Schwinger pair production of electrons and positrons has yet to be experimentally observed, as inaccessibly strong electric fields are required to generate an appreciable rate. High intensity lasers have achieved electric field strengths of $O(0.01\%)$ of the Schwinger critical field strength, $E_c:=\pi m_e^2/g$ \cite{mourou2006optics} and the next generation of high intensity lasers aim to be able to reach electric field strengths of $O(1\%)$ of $E_c$ \cite{mourou2014summary,gales2015implementation,durandeau2018laser}. Though this is not sufficient to observe the original Schwinger process, it may be possible to observe induced Schwinger processes, where the rate is enhanced by some other ingredient. In particular, as we consider in this paper, a thermal bath of photons in the initial state may significantly increase the rate of pair production, and hence one may observe pair production at lower field strengths. One way to experimentally realise a thermal bath of photons would be to use a laser-heated hohlraum, as has been proposed to observe the Breit-Wheeler process, $\gamma\gamma\to e^+e^-$ \cite{breit1934collision,pike2014photon}.

Our initial interest in this calculation was for its relevance to magnetic monopole production in heavy-ion collisions. In the fireball of a heavy-ion collision there are strong magnetic fields and high temperatures. Hence, magnetic monopoles may be produced by the dual thermal Schwinger process \cite{gould2017magnetic}. There is much promise for magnetic monopole searches in current and future heavy-ion collisions, in particular the MoEDAL experiment \cite{moedal2014physics,moedal2018search} at the LHC, so spurring this work. Here we extend the results of Ref. \cite{gould2017thermal} by calculating the prefactor of the rate, an important quantity for making comparison to experiment.

Magnetic monopoles are strongly coupled to the photon. The minimum magnetic charge squared, $g_D^2=(2\pi/e)^2\approx 430$, follows from the Dirac quantisation condition \cite{dirac1931quantised}. Hence one must work to all orders in the coupling to derive results applicable to magnetic monopoles. If one does so and does not consider electric and magnetic charges simultaneously, one can study magnetic monopole Schwinger pair production via its electromagnetic dual. In this case the duality amounts simply to a relabelling of electric degrees of freedom and charges as magnetic. As our calculation reduces to a semiclassical one, we only rely on the classical electromagnetic duality. In the regime considered here, our results are valid both for elementary and 't Hooft-Polyakov monopoles \cite{thooft1974magnetic,polyakov1974particle}.

The paper is set out as follows. In Sec. \ref{sec:sphaleron_rate} we define the sphaleron rate and set up the calculation of it. In Sec. \ref{sec:prefactor_arbitrary_kappa} we find the sphaleron and the spectrum of fluctuations about it. In Sec. \ref{sec:self_force_instability} we note that the spectrum of fluctuations manifests the well-known self-force instability. In Sec. \ref{sec:prefactor_small_kappa} we specialise to a region of parameter space where this instability does not arise. We then calculate the sphaleron rate, including the prefactor, the main result of this paper, which we extend to spin half charged particles in Sec. \ref{sec:prefactor_spinor}. In Sec. \ref{sec:electrons_and_positrons} we discuss the implications of our results for the possible experimental observation of electron-positron pair production from a purely photonic initial state. In Sec. \ref{sec:prefactor_monopoles} we discuss the implications of our results for magnetic monopole searches. In Sec. \ref{sec:prefactor_summary}, we summarise our results.

\section{General approach\label{sec:sphaleron_rate}}

The thermal rate of decay of a metastable state has been studied by many authors \cite{langer1969statistical,affleck1981quantum,linde1981fate,linde1983decay}. In regards to the thermal Schwinger process, in Ref. \cite{gould2017thermal} we calculated the logarithm of the rate, the factor $S$ in $\Gamma(E,T) \sim \mr{e}^{-S}$. For the sphaleron process, we found
\begin{equation}
 \Gamma(E,T)\sim\exp\left[-\frac{2m}{T}\left(1-\sqrt{\frac{g^3E}{4\pi m^2}}\right)\right]. \label{eq:action}
\end{equation}
The difference from the expected Boltzmann suppression in the absence of an external field, $2m/T$, can be understood as a field dependent mass renormalisation. This accounts for the Coulomb corrections to the rest mass of a particle-antiparticle pair at the separation where the external field and Coulomb force balance. The renormalised mass would be $m_*=m(1-\sqrt{g^3E/(4\pi m^2)})$ so that the exponential suppression reads $2m_*/T$. The same kind of mass renormalisation arises in Schwinger pair production at zero temperature \cite{ritus1975lagrange,affleck1981pair,lebedev1984virial}.

To go beyond this, we need an explicit expression for the rate, including the prefactor. As argued for in Refs. \cite{langer1969statistical,affleck1981quantum}, for high temperatures,\footnote{At lower temperatures the rate is given instead by a slightly different expression, related by replacing $|\omega_{-}|\to 2\pi T$ in the prefactor.} where the process is dominated by a static field configuration, a sphaleron, the rate is given by
\begin{equation}
 \Gamma(E,T) \approx \frac{-|\omega_{-}|}{\pi V}\mr{Im}\log(Z),\label{eq:afflecks_rate}
\end{equation}
where $V$ is the spatial volume, $Z$ is the canonical partition function excluding the states containing the decay products and $|\omega_{-}|$ is the rate of growth with time of the unstable mode, responsible for the imaginary part of $\log(Z)$. It is assumed that the rate of decay is slow, so that the process is out of equilibrium.

We first consider the thermal Schwinger process in SQED, with the external field, $E$, pointing along the $x^3$ direction. We write the finite temperature path integral using the imaginary time formalism. The Euclidean Lagrangian for this theory is,
\begin{equation}
 \mathcal{L}_{\mathrm{SQED}}:=\frac{1}{4}F^{\mu \nu}F_{\mu \nu} + D_\mu \phi ( D^\mu \phi)^* + m^2\phi\phi^*, \label{eq:sqedlagrangian}
\end{equation}
where $F_{\mu \nu}=\partial_\mu A_\nu-\partial_\nu A_\mu$ is the field strength and $D_\mu=\partial_\mu+ig A^{\mathrm{ext}}_\mu+ig A_\mu$ is the covariant derivative, showing a split of the photon field into external (or background) and dynamical parts. We assume the scalar self-coupling, i.e. $\lambda(\phi\phi^*)^2/4$, is sufficiently small that we may ignore it, at least in the range of energies considered\textsuperscript{\ref{small_lambda}}. Note that for QED, which we consider later, no such term arises.

The finite temperature partition function for this theory is,
\begin{equation}
Z = \int \mathcal{D}A_\mu\mathcal{D}\phi\ \mathrm{e}^{-\int_x\mathcal{L}_{\mathrm{SQED}}}, \label{eq:Zfields_SQED}
\end{equation}
where the Euclidean time direction has length $1/T$ and the fields satisfy periodic boundary conditions in this direction. We leave the gauge fixing implicit. Eq. \eqref{eq:Zfields_SQED} can be rewritten exactly using the worldline formalism, by carrying out purely formal manipulations \cite{affleck1981pair,gies2005quantum,gould2017thermal}. Essentially a change of integration variables, the usefulness of the worldline representation is that it allows one to circumvent the usual loop expansion, and obtain gauge invariant resummations of (infinite) classes of Feynman diagrams. In Ref. \cite{gould2017thermal} we gave the following exact worldline representation of the thermal partition function in SQED,
\begin{widetext}
\begin{equation}
\frac{1}{V}\mr{Im}\log(Z) = \frac{1}{V}\mr{Im}\log \Bigg[1+
\sum_{n=1}^\infty\frac{1}{n!}\prod\limits_{j=1}^{n}\left( \int_0^\infty \frac{\mr{d}s_j}{s_j}\int\mathcal{D}x_j^\mu \ \mr{e}^{ -\frac{1}{\epsilon}  \Action[x_j;s_j;\kappa,\Temp]} \ \mr{e}^{ \frac{\kappa}{\epsilon} \sum_{k<j} \oint\oint \mr{d}x_j^\mu \mr{d}x_k^\nu G_{\mu\nu}(x_j,x_k;\Temp) }\right)\Bigg], \label{eq:rate_no_approx_prefactor}
\end{equation}
\end{widetext}
where $\epsilon:=gE/m^2$ will act as the semiclassical parameter, akin to $\hbar$. The path integrals over the $x^\mu_j$ are over closed worldlines, with distance measured in units of $m/gE$. In these units the effective coupling between worldlines is $\kappa:=g^3 E/m^2$ and the effective temperature is $\Temp:=mT/gE$. We will refer to $s_j$ as Schwinger parameters. The free thermal photon propagator is $G_{\mu\nu}$, where $\mu,\nu$ run over Euclidean indices $1,2,3,4$. The scaled action, $\Action$, for a single worldline, is\footnote{There are two minor differences here with respect to Ref. \cite{gould2017thermal}. We have rescaled $s\to s/2$ and here we are treating negatively charged particles as matter and positively charged particles as antimatter, which amounts to a plus rather than minus sign for the third term on the right hand side.}
\begin{align}
 \Action&[x;s;\kappa,\Temp] := \frac{s}{2} + \frac{1}{2s}\int^1_0\mr{d}\tau\dot{x}^\mu\dot{x}_{ \mu}+\int_0^1\mr{d}\tau x_{ 3} \dot{x}^4 \nonumber \\
 &- \frac{\kappa}{2}\int_0^1\mr{d}\tau\int_0^1\mr{d}\tau' \dot{x}^\mu(\tau) \dot{x}^\nu(\tau') G_{\mu\nu}(x(\tau),x(\tau');\Temp). \label{eq:scaled_action_s_prefactor}
\end{align} 
The interaction involving the photon propagator has a short distance divergence, corresponding to the electromagnetic contribution to the self-energy of the charged particles. In this paper we regularise this divergence following Polyakov's seminal work \cite{polyakov1980gauge}. We add to the distance squared between points a short distance cut off, $a^2$. Up to gauge fixing terms which vanish upon integration around a closed worldline, the regularised thermal photon propagator is
\begin{align}
&G_{\mu\nu}(x_j,x_k;T;a):= \sum_{n=-\infty}^{\infty}G(x_j,x_k+\frac{n}{T} e_4;a) \delta_{\mu\nu}\nonumber \\
&=\sum_{n=-\infty}^{\infty}\frac{-\delta_{\mu\nu}}{4\pi^2\left((x_j-x_k-\frac{n}{T} e_4)^2+a^2\right)} \nonumber \\
&=\frac{T \sinh \left(2 \pi T \sqrt{r_{jk}^2+a^2}\right)\ \delta_{\mu\nu}}{4 \pi  \sqrt{r_{jk}^2+a^2} \left(\cos \left(2 \pi  T t_{jk}\right)-\cosh \left(2 \pi T \sqrt{r_{jk}^2+a^2}\right)\right)}, \label{eq:thermalsum}
\end{align}
where $e_4$ is the unit vector in the Euclidean time direction and we have defined $t_{jk}:=x_j^4-x_k^4$ and $r_{jk}:=\sqrt{(x_j^1-x_k^1)^2+(x_j^2-x_k^2)^2+(x_j^3-x_k^3)^2}$. For smooth worldlines without self-intersections the only divergence as $a\to 0$ is the self-energy \cite{polyakov1980gauge,dotsenko1979renormalizability,brandt1981renormalization,karanikas1992infrared}, which can be absorbed by adding the following mass counterterm,
\begin{equation}
 -\frac{\kappa }{8 \pi^2}\frac{\pi}{a}\int^1_0\mr{d}\tau\sqrt{\dot{x}^\mu\dot{x}_{ \mu}}. \label{eq:counterterm}
\end{equation}
We may then take the limit $a\to 0$ in the final results. That Eq. \eqref{eq:counterterm} is a mass counterterm can be seen by noting that the first two terms in Eq. \eqref{eq:scaled_action_s_prefactor} are a reparameterisation-fixed form of the same term \cite{polyakov1987gauge}.

Parametrically, the usual loop expansion, given by Eq. \eqref{eq:loop_expansion}, is here mapped to a Taylor expansion in $\kappa$. As mentioned in the introduction, to see the worldline sphaleron requires all orders in the loop expansion, so we do not Taylor expand in $\kappa$. Rather, we note that for $\epsilon\ll 1$ the worldline action, $\Action/\epsilon$, becomes large and a semiclassical approximation should be valid. 

In this context, one would expect the dominant contributions to Eq. \eqref{eq:rate_no_approx_prefactor} to come from configurations consisting of a small number of worldlines. This is because configurations with more worldlines would be expected to have larger actions. Thus, as in Ref. \cite{gould2017thermal}, we perform a cluster expansion of Eq. \eqref{eq:rate_no_approx_prefactor},
\begin{equation}
 \Gamma(E,T) = \sum_{n=1}^{\infty} \Gamma^{(n)}(E,T), \label{eq:cluster_expansion_prefactor}
\end{equation}
where $\Gamma^{(n)}$ is the contribution to $\Gamma$ from clusters of $n$ worldlines. Within the semiclassical approximation, the cluster expansion is a dilute instanton expansion. The leading contribution to the thermal Schwinger rate is given by the instanton with smallest action. This leading order term is approximately equal to the density of these instantons, and is exponentially small. Higher order terms in the cluster expansion are expected to be suppressed by powers of this density, or by subleading instanton densities.

\begin{figure}
 \centering
  \includegraphics[width=1.0\columnwidth]{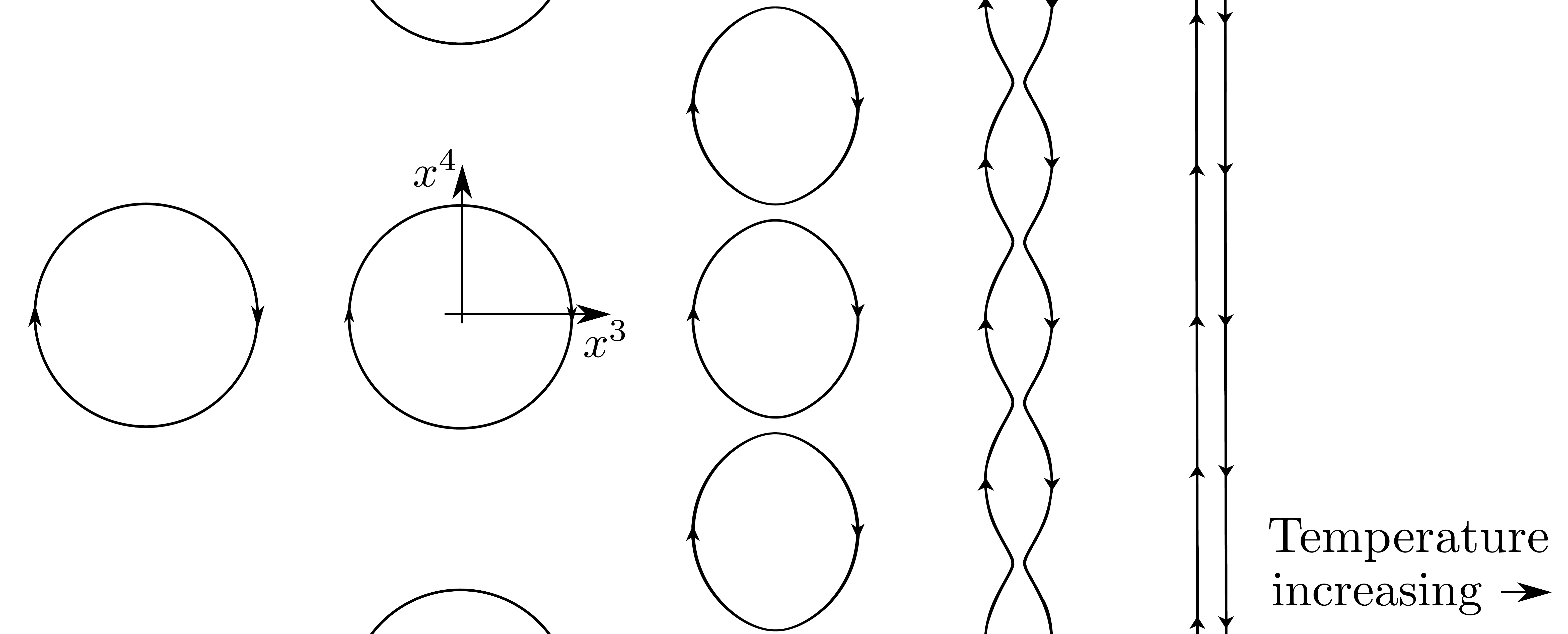}
  \caption[Worldline instantons.]{Worldline instantons relevant for the thermal Schwinger process. Each column represents a different instanton, each relevant for a given temperature. The rightmost instanton is the worldline sphaleron, which dominates the rate at the highest temperatures. The external field points along the 3-direction and the 4-direction is the Euclidean time direction.}
  \label{fig:exampleworldlines}
\end{figure}

At low temperatures, $\Temp\ll 1$, the dominant instanton is the circular worldline instanton \cite{affleck1981pair,affleck1981monopole} (leftmost in Fig. \ref{fig:exampleworldlines}), a saddle point of $\Gamma^{(1)}$. At higher temperatures, thermal corrections deform this instanton, so increasing the rate (second and third from left in Fig. \ref{fig:exampleworldlines}). Above some temperature, $\Temp_{CW}(\kappa)$, where $\Temp_{CW}(0)=1/2$, a second instanton with different topology dominates, called a W instanton in Ref. \cite{gould2017thermal}. It consists of a charged particle and antiparticle oscillating back and forth, parallel to the external field, and is a saddle point of $\Gamma^{(2)}$ (second from right in Fig. \ref{fig:exampleworldlines}). At higher temperatures still, above $\Temp_{WS}(\kappa)$, this W instanton ceases to exist and the dominant instanton is the static worldline sphaleron solution (rightmost in Fig. \ref{fig:exampleworldlines}), also a saddle point of $\Gamma^{(2)}$. The instanton phase diagram outlined here has been established for $0<\kappa\leq 1$ and may be subject to change at larger values of $\kappa$.

\section{The sphaleron and fluctuations about it\label{sec:prefactor_arbitrary_kappa}}

Eq. \eqref{eq:scaled_action_s_prefactor} gives the action for one worldline, $x^\mu(\tau)$, with Schwinger parameter $s$, and it gives the exponent of the integrand of $\Gamma^{(1)}$, when divided by $\epsilon$. In terms of the action for one worldline, the action for two worldlines, the scaled exponent of the integrand of $\Gamma^{(2)}$, is
\begin{align}
 \Action &[x,y;s_x,s_y;\kappa,\Temp] :=  \Action[x;s_x;\kappa,\Temp]+\Action[y;s_y;\kappa,\Temp]\nonumber \\
 &- \kappa\int_0^1\mr{d}\tau\int_0^1\mr{d}\tau' \dot{x}^\mu(\tau) \dot{y}^\nu(\tau') G_{\mu\nu}(x(\tau),y(\tau');\Temp). \label{eq:action_full}
\end{align}
Due to the double integral terms in the action, the corresponding equations of motion are integrodifferential equations. 

The sphaleron is a static solution to these equations of motion. It consists of particle and antiparticle sitting a fixed distance apart along the $x^3$ axis (rightmost in Fig. \ref{fig:exampleworldlines}). It is given by 
\begin{align}
 x(\tau)=x_0(\tau)&:=\bigg\{ 0, 0, \frac{1}{2}\sqrt{\frac{\kappa}{4\pi}}, -\frac{1}{2\Temp}(2\tau -1)\bigg\}, \nonumber \\
 y(\tau)=y_0(\tau)&:=\bigg\{ 0, 0, -\frac{1}{2}\sqrt{\frac{\kappa}{4\pi}}, \frac{1}{2\Temp}(2\tau -1)\bigg\}, \nonumber \\
  s_x=s_y=s_{0}&:=\frac{1}{\Temp}\label{eq:solution_prefactor}
\end{align}
The action of the sphaleron is 
\begin{equation}
 \Action(\kappa,\Temp):=\Action[x_0,y_0;s_{0},s_{0};\kappa,\Temp]=\frac{2}{\Temp}\left(1-\sqrt{\frac{\kappa}{4\pi}}\right).
\end{equation}
Expanding the action to second order about this solution gives a surprisingly large number of terms, $O(100)$, most of which are due to the nonlocal interactions. To proceed we define $\zeta^\mu(\tau):=x^\mu(\tau)-y^\mu(\tau)$ and $\xi^\mu(\tau):=x^\mu(\tau)+y^\mu(\tau)$. The solution given in Eq. \eqref{eq:solution_prefactor} can then be written as
\begin{align}
 \zeta(\tau)=\zeta_0(\tau)&:=\bigg\{ 0, 0, \sqrt{\frac{\kappa}{4\pi}}, -\frac{1}{\Temp}(2\tau -1)\bigg\}, \nonumber \\
 \xi(\tau)=\xi_0(\tau)&:=\bigg\{ 0, 0, 0, 0\bigg\}.
\end{align}
Due to the periodicity, we may expand the fluctuations about the solution in a Fourier series,
\begin{align}
 \zeta^\mu(\tau)-\zeta_0^\mu(\tau)&= \nonumber \\ &a_0^\mu+\sum_{n=1}^{\infty}\left[a_n^\mu\sqrt{2}\cos(2\pi n \tau)+b_n^\mu\sqrt{2}\sin(2\pi n \tau)\right], \nonumber \\
 \xi^\mu(\tau)-\xi_0^\mu(\tau)&=\nonumber \\ &c_0^\mu+\sum_{n=1}^{\infty}\left[c_n^\mu\sqrt{2}\cos(2\pi n \tau)+d_n^\mu\sqrt{2}\sin(2\pi n \tau)\right].
\end{align}
The second order action is diagonal in these Fourier coefficients. It can thus be expressed as
\begin{align}
 \Action^{(2)} =& \frac{1}{2}\Temp (s_x-s_{0})^2 + \frac{1}{2}\Temp (s_y-s_{0})^2 \nonumber \\ &+\frac{1}{2}\sum_{n=0}^{\infty}\sum_{\mu=1}^{4}\left(\alpha_{n}^{\mu}a_n^\mu a_n^\mu+\gamma_{n}^{\mu}c_n^\mu c_n^\mu\right)
 \nonumber \\ &+\frac{1}{2}\sum_{n=1}^{\infty}\sum_{\mu=1}^{4}\left(\beta_{n}^{\mu}b_n^\mu b_n^\mu+\delta_{n}^{\mu}d_n^\mu d_n^\mu\right),\nonumber \\
 =&\frac{1}{2}\Temp (s_x-s_{0})^2 + \frac{1}{2}\Temp (s_y-s_{0})^2 +\frac{1}{2}\sum_i \lambda_i \chi_i \chi_i,\qquad & \label{eq:chi_definition}
\end{align}
where on the last line we have defined $\chi_i$ to run over all the different fluctuations of the worldline, $\{\chi_i\}:=\{a_n^\mu,b_n^\mu,c_n^\mu,d_n^\mu\}$, and $\lambda_i$ to run over all the corresponding eigenvalues, $\{\lambda_i\}:=\{\alpha_n^\mu,\beta_n^\mu,\gamma_n^\mu,\delta_n^\mu\}$. The eigenvalues for $n=0$ are
\begin{align}
 \alpha_{0}&=\{ \frac{2\pi}{\sqrt{\pi \kappa} \Temp}, \frac{2\pi}{\sqrt{\pi \kappa} \Temp}, -\frac{4\pi}{\sqrt{\pi \kappa} \Temp}, 0\}, \nonumber \\
 \gamma_0 &= \{0,0,0,0\}. \label{eq:eigs_0}
\end{align}
The four zero modes of $\gamma_0$ correspond to translations of the instanton. The fifth, $\alpha_{0}^4$, corresponds to translation in the parameter $\tau$. The negative eigenvalue corresponds to increasing, or decreasing, the separation between the particles. It is negative for all $\kappa$ and $\Temp$.

As regards the harmonic modes, by translational symmetry in the Euclidean time direction, one can see that $\gamma_{n} =\beta_n$, and $\delta_{n}=\alpha_{n}$. Further, an $n$th harmonic at a given temperature, $\Temp$, can be seen as $n$ copies of an $n=1$ harmonic at the higher temperature $n\Temp$. Hence, the eigenvalues for $n>1$ are given in terms of the $n=1$ eigenvalues by
\begin{equation}
 \alpha_n(\kappa,\Temp)=n\alpha_1(\kappa,n\Temp), \label{eq:eigs_higher}
\end{equation}
and likewise for the others.
Explicitly the harmonic eigenvalues are found to be,
\begin{align}
 \alpha_{n} = \Bigg\{&\frac{1}{2}(2 \pi n)^2 \Temp-\frac{2}{3} \pi ^2 \kappa  n^3\Temp^2+\frac{\sqrt{\pi }}{\sqrt{\kappa } \Temp}\nonumber \\
&\qquad +\pi\left(  \sqrt{\pi\kappa } n^2\Temp +n+\frac{1}{\sqrt{\pi\kappa}\Temp}\right)e^{-\sqrt{\pi \kappa } n\Temp},\nonumber \\
 &\frac{1}{2}(2 \pi n)^2 \Temp-\frac{2}{3} \pi ^2 \kappa  n^3\Temp^2+\frac{\sqrt{\pi }}{\sqrt{\kappa } \Temp}\nonumber \\
&\qquad +\pi\left(  \sqrt{\pi\kappa } n^2\Temp +n+\frac{1}{\sqrt{\pi\kappa}\Temp}\right)e^{-\sqrt{\pi \kappa } n\Temp},\nonumber \\
   &\frac{1}{2}(2 \pi n)^2 \Temp-\frac{2}{3} \pi ^2 \kappa  n^3\Temp^2-\frac{2
   \sqrt{\pi }}{\sqrt{\kappa } \Temp}\nonumber \\
&\qquad -2 \pi\left(n+\frac{1}{\sqrt{\pi\kappa}\Temp}\right)  e^{-\sqrt{\pi \kappa } n\Temp},\nonumber \\
   &\frac{1}{2}(2\pi n)^2 \Temp\Bigg\}, \label{eq:eigs_alphan}
\end{align}
\begin{align}
 \beta_{n} &= \Bigg\{\frac{1}{2}(2 \pi n)^2 \Temp-\frac{2}{3} \pi ^2 \kappa  n^3\Temp^2+\frac{\sqrt{\pi }}{\sqrt{\kappa } \Temp}\nonumber \\
&\qquad-\pi  \left(\sqrt{\pi\kappa } n^2\Temp+n+\frac{1}{\sqrt{\pi\kappa}\Temp}\right) e^{-\sqrt{\pi \kappa } n\Temp},\nonumber \\
 &\frac{1}{2}(2 \pi n)^2 \Temp-\frac{2}{3} \pi ^2 \kappa  n^3\Temp^2+\frac{\sqrt{\pi }}{\sqrt{\kappa } \Temp}\nonumber \\
&\qquad-\pi  \left(\sqrt{\pi\kappa } n^2\Temp+n+\frac{1}{\sqrt{\pi\kappa}\Temp}\right) e^{-\sqrt{\pi \kappa } n\Temp},\nonumber \\
   &\frac{1}{2}(2 \pi n)^2 \Temp-\frac{2}{3} \pi ^2 \kappa  n^3\Temp^2-\frac{2
   \sqrt{\pi }}{\sqrt{\kappa } \Temp}\nonumber \\
&\qquad+2 \pi\left(n+\frac{1}{\sqrt{\pi\kappa}\Temp}\right)  e^{-\sqrt{\pi \kappa } n\Temp},\nonumber \\
   &\frac{1}{2}(2\pi n)^2 \Temp\Bigg\}. \label{eq:eigs_betan}
\end{align}

\section{Self-force instability\label{sec:self_force_instability}}

\begin{figure}
 \centering
  \includegraphics[width=0.95\columnwidth]{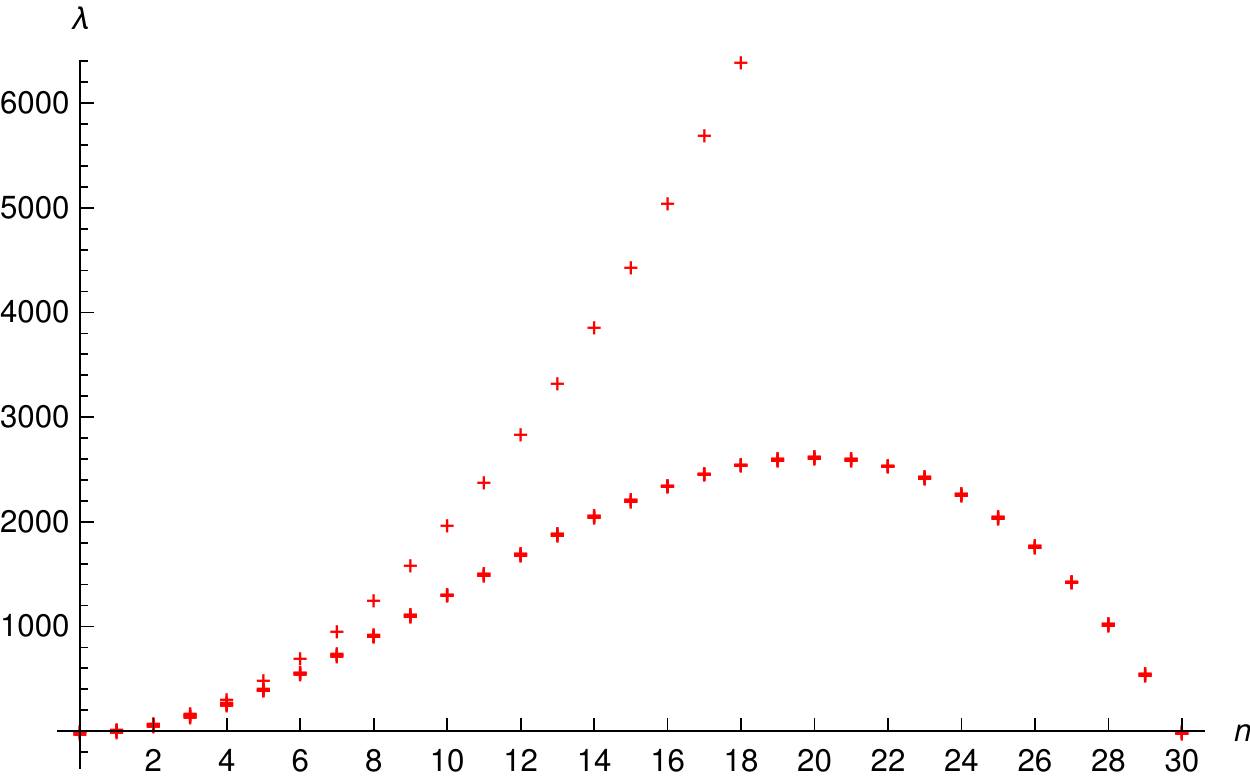}
  \caption[Plot of the first few eigenvalues showing the self-force instability.]{For $(\kappa,\Temp)=(0.1,1.2)$, the first 500 eigenvalues of fluctuations about the sphaleron, ordered by $n$, showing the self-force instability at $n=30$.}
  \label{fig:lowHarmonicEigenvalues}
\end{figure}

These higher harmonics lead to an instability. For sufficiently large $n$, the negative self-force term, $-\tfrac{2}{3} \pi ^2 \kappa \Temp^2 n^3 $, dominates. Hence, there are an infinite number of negative eigenvalues (see Fig. \ref{fig:lowHarmonicEigenvalues}). Within the context of the semiclassical approximation we make, this appears to be a serious problem. On general grounds, one expects any instanton (or sphaleron) to only have a single negative eigenvalue \cite{coleman1987quantum}. However, following Refs. \cite{Weinberg:1992ds,berges1996coarse,strumia1998consistent} we argue that the semiclassical configuration should be a saddle point of the effective action rather than the bare action, with fluctuations above some finite energy scale $\mu$ already integrated out. Then, only fluctuations up to $\mu$ should be included in the fluctuation prefactor. Higher energy fluctuations contribute instead to the renormalisation of parameters. Although we do not explicitly carry out this procedure, we find a regime of parameters where one can separate the scales between the unstable UV modes and the rest of the fluctuations.

The self-force instability is well known in classical electrodynamics \cite{dirac1938classical,wheeler1945interaction,gralla2009rigorous} and in various approximations to quantum electrodynamics \cite{landau1940radius,goebel1970spatial,moniz1974absence,moniz1976radiation,low1997runaway,rosenfelder2004abraham,galley2006electromagnetic}. Its existence in quantum mechanical systems has been linked to the unboundedness of the spectrum of the Hamiltonian \cite{coleman1962runaway,frohlich1973infrared,lieb2010stability}.

In Refs. \cite{moniz1974absence,moniz1976radiation} it was found that a nonrelativistic electron interacting with a quantised photon field does not show the self-force instability, at least for $e^2/(4\pi)\lesssim 1$. Starting from an extended charge distribution, with finite mass, they found that one could take the size of the charge distribution to zero and the result was free of the self-force instability. However, if they changed the orders of the limits and first took the Compton wavelength to zero (or the mass to infinity), the self-force instability was present. Taking the mass to infinity amounts to dropping charged particle loop corrections.

In the dilute instanton gas approximation, extra charged particle loops with $\Action\geq O(1)$ are suppressed by the instanton density, as argued in Sec. \ref{sec:sphaleron_rate}, and Ref. \cite{gould2017thermal}, and hence can justifiably be dropped. On the other hand, extra charged particle loops with vanishing action, as $\epsilon \to 0$, are not present in the semiclassical approximation and are not necessarily suppressed. As $\epsilon\to 0$ there are nontrivial such loops when there is a cancellation between the rest mass and Coulomb interaction terms. These are virtual particles, and they have a size $\sim\kappa/(8\pi)$, or $g^2/(8\pi m)$ in physical units. When there is a separation of scales between the virtual particle loop size, $O(\kappa)$, and the instanton size, $O(1/\Temp)$, i.e. when $\kappa \Temp=g^2T/m\ll 1$, the virtual particle loops should simply renormalise the parameters of the theory \cite{affleck1981pair}, in particular the charge. This is assumed in our analysis. However, when $\kappa\Temp=O(1)$, the saddle-point approximation plus renormalisation is not expected to adequately take virtual charged particle loops into account. The self-force instability may be symptomatic of this.

We define $n_{SF}$ such that the self-force turns those harmonics negative with $n\geq n_{SF}$. For $\Temp=O(1)$ and $\kappa=O(1)$, we find that $n_{SF}=O(1)$. Above $\kappa\approx 3.0653$ and for all $\Temp$, the instability is at $n_{SF}=1$. For $\kappa\Temp\ll 1$, instead we find that 
\begin{equation}
  n_{SF}= \frac{3}{\kappa \Temp}\left[1-\frac{\kappa^{3/2}}{9\pi^{3/2}}+O\left(\kappa^3\right)\right],
\end{equation}
as can be seen in Fig. \ref{fig:lowHarmonicEigenvalues}, where for $(\kappa,\Temp)=(0.1,1.2)$ we can see that $n_{SF}=30$. For $\kappa\Temp\ll 1$ the self-force problem is moved to parametrically high harmonics, or short distances, where the effects of virtual charged particle pairs are significant and hence the naive semiclassical approximation is expected to break down. For the purposes of the semiclassical calculation, one should cut off the higher harmonics below $O(1/(\kappa\Temp))$. The self-force instability is thereby moved into the ultraviolet, where it can be considered separately, and should contribute only to the renormalisation of couplings.

For electric charges the coupling is weak, $g^2:=e^2\ll 1$. In this case $\kappa\ll\epsilon\ll 1$, and the self-force problem is not present at leading order in $\epsilon$, which is the approximation we make.
For magnetic charges, on the other hand, $g_M\gg 1$ and hence $\kappa\gg \epsilon$. Magnetic and electric charges, $g_M$ and $e$, are related inversely via the Dirac quantisation condition, $g_M e=2\pi j$, where $j\in \mathbb{Z}$. This relationship is expected to hold for the running couplings \cite{calucci1982renormalization,coleman1982magnetic,goebel1983antishielding}. Hence, as one probes shorter distances the effective magnetic charge decreases; magnetic charge is anti-shielded. This has been argued to be an effective spreading-out of magnetic charge over scales $O(g_M^2/(4\pi m))$ \cite{goebel1970spatial}. Classically, a charge distribution spread out on these scales is stable \cite{moniz1974absence,moniz1976radiation}, hence, for magnetic monopoles, one would expect that the ultraviolet physics does not suffer from the self-force instability.

\section{The scalar prefactor}\label{sec:prefactor_small_kappa}

In this section we consider the regime for which the self-force instability is not present. Thus we assume that $\kappa\Temp \ll 1$ as discussed in the previous section. We will further assume that $\kappa\Temp^2 \ll 1$, in which case the eigenvalues, $\lambda_i$, of Eqs. \eqref{eq:eigs_0}, \eqref{eq:eigs_alphan} and \eqref{eq:eigs_betan}, simplify considerably. They are given in Table \ref{table:eigenvales}. We assume the running coupling is known at the relevant energy scale.

\begin{table}
\begin{center}
\begin{tabular}{ c c } 
  \toprule
  $\lambda_i$ & Multiplicity \\
  \colrule
  $0$ & 5 \\
  $\frac{\sqrt{4\pi}}{\sqrt{\kappa}\Temp}$ & 2 \\
  $-\frac{2\sqrt{4\pi}}{\sqrt{\kappa}\Temp}$ & 1 \\[1ex]
  \colrule 
  $\frac{1}{2}(2\pi n)^2\Temp$ & 10 \\
  $\frac{1}{2}\left((2\pi n)^2\Temp+\frac{2\sqrt{4\pi}}{\sqrt{\kappa}\Temp}\right)$ & 4 \\
  $\frac{1}{2}\left((2\pi n)^2\Temp-\frac{4\sqrt{4\pi}}{\sqrt{\kappa}\Temp}\right)$ & 2 \\
  \botrule
\end{tabular}
\caption[Table of fluctuation eigenvalues.]{Table of eigenvalues of the fluctuations about the worldline sphaleron assuming $\kappa\Temp^2\ll 1$. The lower single line separates constant fluctuations from harmonic ones.}
\label{table:eigenvales}
\end{center}
\end{table}
\begin{table}
\begin{center}
\begin{tabular}{ c c } 
  \toprule
  $\lambda_{0i}$ & Multiplicity \\
  \colrule
  $0$ & 8 \\
  \colrule
  $(2\pi n)^2\Temp$ & 16 \\
  \botrule
\end{tabular}
\caption[Table of free particle eigenvalues.]{Table of eigenvalues of the square of the corresponding free particle path integral. The lower single line separates constant fluctuations from harmonic ones.}
\label{table:eigenvalues0}
\end{center}
\end{table}

The temperature, $\Temp_{WS}$, above which the worldline sphaleron dominates the thermal Schwinger process satisfies $\Temp_{WS}\gtrsim 0.5$, at least for $\kappa\leq 1$. For small $\kappa$, it grows and is given approximately by
\begin{equation}
 \Temp_{WS}\approx \frac{\sqrt{2}}{\pi^{3/4}\kappa^{1/4}}. \label{eq:approx_tws}
\end{equation}
Given that $\Temp>\Temp_{WS}$ and $\kappa\Temp^2\ll 1$, we must have that $\kappa\ll 1$ and $\Temp$ must lie in the window
\begin{equation}
 \frac{\sqrt{2}}{\pi^{3/4}\kappa^{1/4}}<\Temp \ll \frac{1}{\sqrt{\kappa}}.
\end{equation}

Above $\Temp_{WS}$, the rate is dominated by the sphaleron and hence by the second term in the cluster expansion,
\begin{align}
 \Gamma(E,T) &\approx \Gamma^{(2)}(E,T), \nonumber \\
 \Gamma^{(2)}(E,T)&=-\frac{|\omega_-|}{\pi V}\mr{Im}\frac{1}{2!}\int_0^\infty \frac{\mr{d}s_x}{s_x}\int_0^\infty \frac{\mr{d}s_y}{s_y} \  \nonumber \\ & \int\mathcal{D}x^\mu\int\mathcal{D}y^\mu\mr{e}^{ -\frac{1}{\epsilon}  \Action[x,y;s_x,s_y;\kappa,\Temp] }.
\end{align}
We wish to evaluate this in the saddle point approximation about the sphaleron. In this approximation, it is given by
\begin{align}
 \Gamma(E,T) \approx &-\frac{|\omega_-|}{\pi V}\Temp^2\mr{e}^{ -\frac{1}{\epsilon}\Action(\kappa,\Temp)}\mr{Im} \int_{-\infty}^\infty \mr{d}s_x \mr{e}^{-\frac{\Temp}{2\epsilon}(s_x
-s_0)^2}\nonumber \\ &\int_{-\infty}^\infty \mr{d}s_y \mr{e}^{-\frac{\Temp}{2\epsilon}(s_y-s_0)^2}\int\mathcal{D}\chi_i \ \mr{e}^{ -\frac{1}{2\epsilon}  \sum_i\lambda_i\chi_i\chi_i },\nonumber \\
\approx -&\frac{2\epsilon\Temp |\omega_-|}{V}\mr{e}^{ -\frac{1}{\epsilon}\Action(\kappa,\Temp)}\ \mr{Im} \int\mathcal{D}\chi_i \ \mr{e}^{ -\frac{1}{2\epsilon}  \sum_i\lambda_i\chi_i\chi_i },\label{eq:z_saddle}
\end{align}
where $\chi_i$ and $\lambda_i$ are defined in Eq. \eqref{eq:chi_definition}.

We divide the integrations up into the constant fluctuations, which correspond simply to translations of the worldlines, and the harmonic fluctuations, which correspond to sines and cosines. The final result involves a product of the contribution from the constant fluctuations, $\mc{C}$, and the harmonic fluctuations, $\mc{H}$,
\begin{equation}
 \int\mathcal{D}\chi_i \ \mr{e}^{ -\frac{1}{2\epsilon}  \sum_i\lambda_i\chi_i\chi_i }=:\mc{C}\ \mc{H}.
\end{equation}
We consider $\mc{C}$ first. From Table \ref{table:eigenvales}, we can see that there are eight constant fluctuations: five of these are zero modes, one is a negative mode and the remaining two are positive. Again, the result involves a product of the contribution from these three groups,
\begin{equation}
 \mc{C}=:\mc{C}_Z\ \mc{C}_N\ \mc{C}_P.
\end{equation}
To perform the integrations over the zero modes, we first put the worldlines in a large box with spatial sides of length $\tilde{L}$ (in units of $m/gE$), and then in the result drop terms subdominant in $\tilde{L}$. Pairing up the spatial zero modes with the nonzero constant modes, we make use of the following elementary integral,
\begin{align}
 \int_{-\tilde{L}/2}^{\tilde{L}/2}\mr{d}x&\int_{-\tilde{L}/2}^{\tilde{L}/2}\mr{d}y\ \mr{e}^{-\frac{\lambda}{2\epsilon}(x-y)^2}=\nonumber \\
 &\tilde{L}\int_{-\infty}^{\infty}\mr{d}\zeta\ \mr{e}^{-\frac{\lambda}{2\epsilon}\zeta^2}\left[1+O\left(\frac{\sqrt{\epsilon}}{\sqrt{\lambda}\tilde{L}}\right)\right].
\end{align}
Using this, and doing the trivial integrals over the two zero modes in the 4 direction, one can find that 
\begin{equation}
 \mc{C}_Z=\frac{\tilde{L}^3}{\tilde{T}^2}=\frac{m^3\epsilon^3 V}{\tilde{T}^2},
\end{equation}
where $V$ is the spatial volume in standard dimensionful units.

Defining the integration over the negative mode requires an analytic continuation. This is done following the classic work of Langer \cite{langer1967theory}, resulting in an overall factor of $1/2$ on top of the naive result,
\begin{equation}
 \mc{C}_N=\frac{1}{2}(2\pi)^{1/2}\left(-\frac{2\sqrt{4\pi}}{\epsilon \sqrt{\kappa}\Temp }\right)^{-1/2}=\pm i\frac{1}{2\sqrt{2}} (\pi\kappa)^{1/4} \sqrt{\Temp \epsilon }.
\end{equation}
The sign ambiguity arises in the process of analytic continuation and we must choose the negative sign. The integrations over the two positive modes are elementary,
\begin{equation}
 \mc{C}_P= (2\pi)^{2/2}\left(\frac{\sqrt{4\pi}}{\epsilon \sqrt{\kappa}\Temp }\right)^{-2/2}= \sqrt{\pi \kappa } \Temp \epsilon.
\end{equation}
 Thus we arrive at
\begin{equation}
 \mc{C}=\frac{-i \pi ^{3/4} \kappa ^{3/4} m^3 \epsilon ^{9/2} V}{2 \sqrt{2} \sqrt{\Temp}}.
\end{equation}

To perform the infinite integrations over the harmonic fluctuations requires regularisation or, equivalently, normalisation. We normalise the path integral measure with respect to the square of the equivalent free particle path integral,
\begin{align}
 \left(\int\mathcal{D}x^\mu \mr{e}^{-\frac{\Temp}{2\epsilon}\int_0^1\dot{x}^2\mr{d}\tau}\right)^2&=\int\mathcal{D}\chi_i \mr{e}^{-\frac{1}{2\epsilon}\sum_i\lambda_{0i}\chi_i\chi_i},\nonumber \\
 =&\frac{\tilde{L}^6}{\tilde{T}^2}\int\mathcal{D}'\chi_i \mr{e}^{-\frac{1}{2\epsilon}\sum_i'\lambda_{0i}\chi_i\chi_i}, \label{eq:normalisation_denominator} \\
 =&\frac{\tilde{L}^6}{\tilde{T}^2}\frac{\Temp^4}{(2\pi \epsilon)^4},\label{eq:normalisation_numerator}
\end{align}
where the $\lambda_{0i}$ are defined by this equation, and are given in Table \ref{table:eigenvalues0}. The final result is the usual one for (eight powers of) the free particle path integral in 1D quantum mechanics \cite{feynman2010quantum} with $ i m/\mc{T} \to -\Temp/\epsilon$ where $\mc{T}$ refers to time elapsed and the boundary conditions on the path integrals are periodic. In the second line we have factored off the contribution from the constant modes. The $'$ on the summation symbol and in the integration measure denotes that only the harmonic modes are included.

In the integrations over the harmonic modes we must keep in mind that the change of variables, $(x^\mu(\tau),y^\mu(\tau))\to(\zeta^\mu(\tau)=x^\mu(\tau)-y^\mu(\tau),\xi^\mu(\tau)=x^\mu(\tau)+y^\mu(\tau))$, was carried out. The Jacobian of the transformation is $1/2$ for each pair of degrees of freedom or $1/\sqrt{2}$ for each degree of freedom. This can be seen easily in the two dimensional transformation $(x,y)\to (\zeta=x-y,\xi=x+y)$,
\begin{equation}
 |J| =
 \left| \begin{array}{cc}
\frac{\partial x}{\partial \zeta} & \frac{\partial x}{\partial \xi}  \\
\frac{\partial y}{\partial \zeta} &\frac{\partial y}{\partial \xi}  
\end{array} \right|
= 
\left| \begin{array}{cc}
\frac{1}{2} & \frac{1}{2}  \\
-\frac{1}{2} &\frac{1}{2}  
\end{array} \right|
=\frac{1}{2}. 
\end{equation}
Multiplication by this Jacobian factor is equivalent to multiplying the eigenvalues by $2$.

Multiplying by Eq. \eqref{eq:normalisation_numerator} and dividing by Eq. \eqref{eq:normalisation_denominator}, $\mc{H}$ then takes the form
\begin{equation}
 \mc{H}=\frac{\Temp^4}{(2\pi \epsilon)^4}\prod_i'\left(\frac{2\lambda_i}{\lambda_{0i}}\right)^{-1/2},
\end{equation}
where again the $'$ on the product symbol denotes that only the eigenvalues corresponding to harmonic modes are included. The infinite products of the ratios of these eigenvalues are well defined and can be evaluated using the following identity,
\begin{equation}
 \prod_{n=1}^{\infty}\left(1-\frac{c^2}{n^2}\right)^{-1}=\frac{\pi c}{\sin\left(\pi c\right)}. \label{eq:product_identities}
\end{equation}
The result is
\begin{equation}
 \mc{H}=\frac{\Temp  }{8
   \sqrt{2} \pi ^{13/4} \kappa ^{3/4} \epsilon ^4\sin \left(\frac{\sqrt{2} \pi^{1/4}}{\kappa^{1/4}  \Temp}\right)\text{sinh}^2\left(\frac{\pi^{1/4}}{\kappa^{1/4} \Temp}\right)}.
\end{equation}

Putting it all together we find
\begin{equation}
 \frac{1}{V}\mr{Im}\log(Z)\approx -\frac{m^3 (\Temp \epsilon) ^{3/2} \mr{e}^{-\frac{2}{\Temp\epsilon}\left(1-\sqrt{\frac{\kappa}{4\pi}}\right)} }{16 \pi^{3/2}\sin \left(\frac{\pi  \Temp_{WS}}{\Temp}\right)\sinh^2\left(\frac{\pi  \Temp_{WS}}{\sqrt{2} \Temp}\right)} ,\label{eq:logZ}
\end{equation}
where we have reintroduced $\Temp_{WS}\approx\sqrt{2}/(\pi^{3/4}\kappa^{1/4})$ to simplify the expression. Note that it is consistent to keep the $\sim \sqrt{\kappa}$ term in the exponent, though such terms were dropped in the eigenvalues, as the correction is multiplicative rather than additive, as well as coming with a negative power of $\epsilon$.

Eq. \eqref{eq:logZ} is divergent at $\Temp=\Temp_{WS}$, where the sphaleron develops another zero mode. Below this temperature the sphaleron is unstable due to the existence of another saddle point with lower action, the W instanton of Ref. \cite{gould2017thermal}. This may signal a parametric enhancement of the dependence on the semiclassical parameter, $\epsilon$ \cite{levkov2007unstable}. This can be investigated by expanding the action to higher than second order in the harmonic fluctuation which becomes a zero mode at $T=T_{WS}$. Then one may perform the non-Gaussian integral over this mode to find the correction to the prefactor (see for instance \cite{ivlev1987tunneling}). Whether or not there are phenomenological consequences of this enhancement depends on precise values in a given situation, though the exponential dependence may well dominate over such power law enhancements in the prefactor. The higher harmonic divergences, at $\Temp=\Temp_{WS}/n$, are not relevant as they exist at lower temperatures where the sphaleron does not dominate the rate.

The last ingredient required to construct the sphaleron rate is $|\omega_-|$, the rate of growth with time of the unstable mode. This is
\begin{equation}
 |\omega_-| = 2\pi T_{WS} \approx 2\pi \left(\frac{4 g E^3}{\pi^3 m^2}\right)^{1/4}.
\end{equation}

\begin{figure}
 \centering
  \includegraphics[width=0.95\columnwidth]{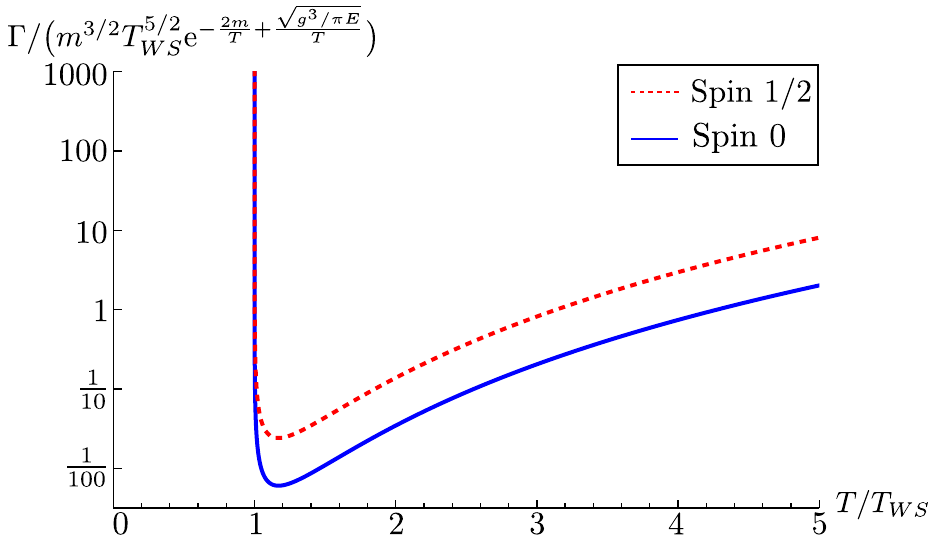}
  \caption[Plot of the sphaleron rate prefactor.]{The sphaleron rate prefactor. The sharp rise at $T=T_{WS}$ is a divergence of the form $c/(T-T_{WS})$, where $c$ is a constant. It is due to the presence of an extra zero mode.}
  \label{fig:prefactor}
\end{figure}

The rate of pair production of scalar charged particles is thus given by
\begin{align}
 \Gamma(E,T)\approx &\frac{T_{WS}\left(mT\right)^{3/2}\mr{e}^{-\frac{2m}{T}+\frac{\sqrt{g^3 E/\pi}}{ T}}}{(4\pi)^{3/2}\sin \left(\frac{\pi T_{WS} }{T }\right)\sinh^2\left(\frac{\pi T_{WS}}{\sqrt{2} T }\right)} \nonumber \\ 
 &\quad\left[1+O\left(\frac{g E}{m^2},\frac{g^3 E}{m^2},\frac{T}{m},\frac{g^2T}{m},\frac{gT^2}{E}\right)\right], \label{eq:rate_langer_entire}
\end{align}
where we have restored the dimensionful variables and the result is valid for temperatures satisfying
\begin{equation}
 T>T_{WS}\approx\left(4 g E^3/\pi^3 m^2\right)^{1/4}.
\end{equation}
The prefactor is plotted in Fig. \ref{fig:prefactor}.

\section{The spinor prefactor}\label{sec:prefactor_spinor}

For QED with (Dirac) spinor charged particles, just as for SQED, one can formally represent the exact partition function in terms of an infinite sum of integrals over worldlines. Compared to SQED each worldline path integral contains an additional spin-dependent factor. This can be seen by starting from the following identities
\begin{align}
 \mr{det}(\slashed{D}+ m)&=\mr{det}(-\slashed{D}+ m),\nonumber \\
 =\mr{det}&\left(-\slashed{D}^2+ m^2\right)^{1/2},\nonumber \\
 =\mr{det}&\left(-D^2+ m^2-\frac{i}{2}g\Sigma^{\mu\nu}(F^{ext}_{\mu\nu}(x)+F_{\mu\nu}(x))\right)^{1/2},\label{eq:det_fermion}
\end{align}
where the $\Sigma^{\mu\nu}$ are proportional to the generators of Lorentz transformations in the spin 1/2 representation, i.e. $\Sigma^{\mu\nu}=[\gamma^\mu,\gamma^\nu]/2$, where $\gamma^\mu$ are the Euclidean gamma matrices (see Ref. \cite{laine2016basics} for a definition). Using the final line of Eq. \eqref{eq:det_fermion}, one can see that the spin-1/2 functional trace can be written in terms of the spin-0 trace and an additional spin factor \cite{feynman1951operator,polyakov1988two,corradini2015spinning}
\begin{align}
 \mr{Tr} (\mathrm{e}^{-(\slashed{D}+ m)s}) &= \mathrm{Tr} (\mathrm{e}^{-(-D^2+ m^2-\frac{i}{2}g\Sigma^{\mu\nu}(F^{ext}_{\mu\nu}(x)+F_{\mu\nu}(x))s})\nonumber \\
 =\int \mathcal{D}x^\mu &\mathrm{e}^{-S_0[x^\mu,A^{ext}_\mu+A_\mu;s]}Spin[x^\mu,A^{ext}_\mu+A_\mu;s],
\end{align}
where $S_0[x^\mu,A^{ext}_\mu+A_\mu;s]$ is what one would get for a spin zero particle and the spin factor is given by
\begin{equation}
 Spin[x^\mu,A^{ext}_\mu+A_\mu;s]:= Tr_\gamma \mathcal{P} \ \mathrm{e}^{ i \frac{g}{2} \int_0^s \mathrm{d}\tau \Sigma^{\mu\nu} (F^{ext}_{\mu\nu}(x)+F_{\mu\nu}(x))} \label{eq:spin_factor}
\end{equation}
where $ Tr_\gamma $ signifies the trace over spinorial indices and $\mathcal{P}$ is the path ordering operator.

After scaling $\tau\to \tau/s$, $s\to s/(gE)$ and $x\to (m/gE)x$, and performing the Gaussian integrations over the dynamical gauge field, one can show that the spin factor is subleading in $\epsilon$ versus the spinless part of the action (see Appendix A of Ref. \cite{gould2017thermal}). The spin factor does however modify the prefactor at leading order in $\epsilon$. To this order we need only to evaluate the spin factor on the worldline sphaleron. This gives, for each worldline, $x$, a multiplicative factor of
\begin{widetext}
\begin{align}
 -\frac{1}{2}\mr{Tr}_{\gamma_x}\mc{P}_x\exp\bigg[  s_{0} \Sigma^{34}_{x} + &\kappa s_{0} \int_0^1\mr{d}\tau \int_0^1\mr{d}\tau' \Sigma_{x}^{\mu\rho}\left(\partial^{x_0}_\rho G_{\mu\nu}(x_0,y_0';\Temp)\dot{y_0}^{'\nu}+\partial^{x_0}_\rho G_{\mu\nu}(x_0,x_0';\Temp)\dot{x_0}^{'\nu}\right)\bigg] \nonumber \\
 &=-\frac{1}{2}\mr{Tr}_{\gamma_{x}}\mc{P}_{x}\exp\left[ \Sigma_x^{34}\left(\frac{1}{2\Temp}-\frac{1}{2\Temp}\right)\right],\nonumber \\
 &=-2,
\end{align}
\end{widetext}
where $\mr{Tr}_{\gamma_x}$ denotes a trace over spinor indices of the ${x}$ worldline, $\mc{P}_x$ is the path ordering operator along the $x$ worldline, $\Sigma^{\mu\nu}:=[\gamma^\mu,\gamma^\nu]/2$, where $\gamma^\mu$ are the Euclidean Dirac (gamma) matrices satisfying $\{\gamma^\mu,\gamma^\nu\}=2\delta^{\mu\nu}$ and $'$ denotes dependence on $\tau'$ as opposed to $\tau$. There is a second factor, for the other worldline, $y$, which differs by interchange of $x$ and $y$. The two factors are equal to each other. Hence the worldline sphaleron rate for spinor charged particles is
\begin{equation}
 \Gamma_{\mr{spinor}}(E,T)\approx 4\Gamma_{\mr{scalar}}(E,T)\label{eq:rate_langer_spin_half}.
\end{equation}
The 4 comes from a product of traces over gamma matrices and hence can be seen as a factor of $(2s+1)^2$, where $s$ is the spin, to be compared with the single power of $(2s+1)$ which arises in Schwinger pair production at zero temperature. Thus spin enters the rate of pair production essentially trivially. The factor of $(2s+1)^2$ simply counts the possible spins of the produced particles: up-up, up-down, down-up and down-down. This result can be seen as a natural extension of the factor $(2s+1)$ which arises in Schwinger pair production. In that case, in vacuum, the produced pairs must have opposite spins so only up-down and down-up are possible. However, in our case, angular momentum can be exchanged with the thermal bath of photons.

Note that the effects of spin and statistics should give non-trivial corrections in stronger fields, at higher temperatures, or simply at later times, where the back reaction of the produced charged particles cannot be ignored. However, in the physical situation we consider neither Pauli blocking nor Bose enhancement are relevant because our initial state contains no charged particles and, in the presence of the weak external field, particles are only produced exponentially slowly.

Combining this result with that of the previous section, we arrive at the chief result of this paper, Eq. \eqref{eq:rate_intro}.

\section{Electrons and positrons} \label{sec:electrons_and_positrons}

So, what do our calculations imply for the pair production of the lightest electric particles in nature, the electron and the positron? In this case, the theory is weakly coupled, $g^2:=e^2\ll 1$, where $e$ is the charge of a positron, and hence we may take the weak coupling limit of Eq. \eqref{eq:rate_intro}, in which some of the approximations become redundant and the equation becomes
\begin{align}
 \Gamma(E,T)\approx &\frac{4 T_{WS}\left(m_eT\right)^{3/2}\mr{e}^{-\frac{2m_e}{T}+\frac{\sqrt{e^3E/\pi}}{T}}}{(4\pi)^{3/2}\sin \left(\frac{\pi T_{WS} }{T }\right)\sinh^2\left(\frac{\pi T_{WS}}{\sqrt{2} T }\right)} \nonumber \\
 &\quad\quad\quad \left[1+O\left(e^2,\frac{e E}{m_e^2},\frac{T}{m_e},\frac{e T^2}{E}\right)\right],\label{eq:rate_intro_electric}
\end{align}
where $m_e$ is the mass of the electron and $T_{WS}\approx\left(4 e E^3/\pi^3 m_e^2\right)^{1/4}$. Note that there are now only four dimensionless combinations of parameters in the big $O$, rather than five. We describe as the region of validity of this expression, where all of these dimensionless parameters are small, and where $T>T_{WS}$.

It is instructive to compare Eq. \eqref{eq:rate_intro_electric} to known results for $E=0$ and for $T=0$. In the former case, in a thermal bath of photons with zero electric field, to leading order, electron-positron pair production proceeds via the collision of pairs of photons, $\gamma\gamma\to e^+e^-$, the Breit-Wheeler process \cite{breit1934collision}. Above the kinematic threshold this process has cross section approximately equal to the classical size of the electron, $\sigma_{\mr{BW}} \sim e^4/(16\pi^2 m_e^2)$. Averaging this over the distribution of energies present in a thermal bath of photons one finds \cite{king2012pair},
\begin{equation}
 \Gamma_{\mr{BW}}(T)\approx \frac{e^4 m_e T^3}{2(2\pi)^4}\mr{e}^{-\frac{2m_e}{T}}\left[1+O\left(e^2,\frac{T}{m}\right)\right].\label{eq:breit_wheeler}
\end{equation}
Adding a constant electric field does not change the rate of this process, as the additional photons have infinite wavelength and hence zero energy.

The exponential dependences of Eqs. \eqref{eq:rate_langer_spin_half} and \eqref{eq:breit_wheeler} are similar, though the prefactors differ markedly. In the presence of an electric field and a thermal bath of photons with $T> T_{WS}$, both processes are possible. In this case the ratio of the worldline sphaleron rate, $\Gamma(E,T)$, and the Breit-Wheeler rate is
\begin{align}
\frac{\Gamma(E,T)}{\Gamma_{\mr{BW}}(T)}&\approx\frac{\pi}{4}\left(\frac{4\pi}{e^2}\right)^{-15/8} \left(\frac{m_e^2}{e E}\right)^{3/8}  F\left(\frac{T}{T_{WS}}\right)\mr{e}^{\frac{\sqrt{e^3E/\pi}}{T}}, \nonumber \\
&\gtrsim 1754 \left(\frac{m_e^2}{E}\right)^{3/8}\mr{e}^{\frac{\sqrt{e^3E/\pi}}{T}}\gg 1,
\end{align}
where $F(t)$ is defined by this equation and the second line follows by evaluating $F(t)$ at its minimum, at $t\approx 1.37$. Hence, in the domain of validity of both results, the worldline sphaleron rate is faster than the (purely thermal) Breit-Wheeler rate for all nonzero electric fields. In practice, the most easily physically realisable regime, is probably the regime of weakest fields, $E\ll \pi(m^2 T^4/4e)^{1/3}$ (equivalent to $T\gg T_{WS}$), where the ratio of rates reduces to
\begin{equation}
\frac{\Gamma(E,T)}{\Gamma_{\mr{BW}}(T)}\approx \frac{16 \pi  (m T)^{3/2}}{e^{9/2} E^{3/2}}\mr{e}^{\frac{\sqrt{e^3E/\pi}}{T}}\gg 1.
\end{equation}
In the regime of validity this amounts to a factor $O(10^6)$. The surprising result is that for weak fields, $E\ll m_e^2$, the worldline sphaleron rate is parametrically faster than the Breit-Wheeler rate, despite the fact one would expect the former to reduce to the latter as $E\to 0$. However, one cannot take the $E\to 0$ limit at fixed $T$ while staying in the region of validity of our calculation, due to the requirement $E\gg eT^2$ (see Eq. \eqref{eq:rate_intro_electric}).

We can also compare the worldline sphaleron rate with that of the Schwinger rate at zero temperature \cite{schwinger1951gauge},
\begin{equation}
 \Gamma_{\mr{Schwinger}}(E)\approx\frac{(e E)^2}{4\pi^3}\mr{e}^{-\frac{\pi m_e^2}{eE}}\left[1+O\left(e^2,\frac{e E}{m^2}\right)\right].
\end{equation}
In this case, in the regime $T>T_{WS}$, and for weak fields, $E\ll m_e^2$, using the same reasoning as for the comparison with the Breit-Wheeler rate, we find,
\begin{equation}
 \frac{\Gamma(E,T)}{\Gamma_{\mr{Schwinger}}(E)}\gtrsim 5.81 \left(\frac{m^2}{E}\right)^{1/8} \mr{e}^{\frac{\pi m_e^2}{eE}-\frac{2m_e}{T}+\frac{\sqrt{e^3E/\pi}}{T}}\gg 1.
\end{equation}
In the region of validity, the worldline sphaleron rate is exponentially faster than the zero temperature Schwinger rate. Numerically, one finds the enhancement to be in the region of $10^{10^3}$.

\section{Magnetic Monopoles\label{sec:prefactor_monopoles}}

As mentioned in the introduction, our initial interest in this calculation was for its relevance to magnetic monopole pair production, and hence to searches for magnetic monopoles. Due to the necessary strong coupling of magnetic monopoles, $g_M^2\geq g_D^2:=(2\pi/e)^2\gg 1$, perturbative techniques for computing their production cross sections, $\sigma_{ab\to M\bar{M}}$, fail. As a consequence, these cross sections are largely unknown and, in fact, nonrigorous arguments for the order of magnitude of the cross section differ by hundreds of orders of magnitude. This huge discrepancy is due to the presence or absence of an exponential suppression of the form $\exp(-16\pi/e^2)$. For monopole production in collisions of ``small" particles, such as $pp\to M\bar{M}$ or $e^+e^-\to M\bar{M}$, this exponential suppression has been argued to be present for 't Hooft-Polyakov monopoles \cite{witten1979baryons,drukier1982monopole} and may also be argued to be present for elementary (Dirac) monopoles \cite{goebel1970spatial}. 

Experimental collider searches for magnetic monopoles have largely focused on collisions of ``small" particles, such as protons or electrons \cite{tanabashi2018review}. The null results of these searches have led to upper bounds on cross sections for the pair production of magnetic monopoles. Unfortunately, given the huge theoretical uncertainties on these cross sections, such experimental bounds cannot be used to derive bounds on the properties of any possible monopoles, such as their mass, $m_M$.

This problem does not extend to heavy ion collisions. In this case there are strong magnetic fields \cite{kharzeev2007effects,huang2015electromagnetic} and high temperatures \cite{baier2000bottom,alice2015direct} and thus magnetic monopoles may be produced by the dual of the thermal Schwinger process. From a consideration of this, one can explicitly calculate the cross section for magnetic monopole pair production. This has led to the first (lower) bounds on the mass of possible magnetic monopoles from any collider search \cite{gould2017magnetic}. These are also the strongest reliable bounds on $m_M$, though they are surprisingly weak, being only $O$(GeV). One can expect to see significant improvements in the near future, as the heavy-ion collisions which gave rise to that bound \cite{he1997search} had a centre of mass energy per nucleon of only 8.7GeV, much lower than, for example, the last lead-lead run at the LHC in 2015, which had a centre of mass energy per nucleon of 5020GeV.


As briefly argued in the introduction, the rate of thermal Schwinger pair production that we have calculated in QED at arbitrary coupling, is directly applicable to magnetic monopole production. This is due to classical electromagnetic duality, a symmetry of Maxwell's equations extended to include magnetic charges,
\begin{align}
F^{\mu\nu} \to *F^{\mu\nu}, &\qquad *F^{\mu\nu} \to -F^{\mu\nu}, \nonumber \\
j_E^\mu \to j_M^\mu, &\qquad j_M^\mu \to -j_E^\mu,
\end{align}
where $*$ denotes the Hodge dual, $j_E^\mu$ denotes the electric current and $j_M$ the magnetic current. In our result, Eq. \eqref{eq:rate_intro}, this amounts simply to a relabelling of electric quantities as magnetic. Under this duality our result is directly applicable to elementary (Dirac) monopoles. For 't Hooft-Polyakov monopoles, which are composite, solitonic objects, our worldline calculation is also applicable because their size is $O(\kappa/4\pi)$ in our scaled units and hence is parametrically smaller than the worldline sphaleron. Thus an effective description of 't Hooft-Polyakov monopoles as magnetically-charged worldlines is applicable \cite{bardakci1978local,manton1978effective,affleck1981monopole,gould2017thermal}.

Considering that magnetic monopoles are necessarily strongly coupled, we may take the strong coupling limit of Eq. \eqref{eq:rate_intro}, in which some other of the approximations become redundant and the equation becomes
\begin{align}
 \Gamma(B,T)\approx &\frac{(2s+1)^2 T_{WS}\left(m_M T\right)^{3/2}\mr{e}^{-\frac{2m_M}{T}+\frac{\sqrt{g_M^3B/\pi}}{T}}}{(4\pi)^{3/2}\sin \left(\frac{\pi T_{WS} }{T }\right)\sinh^2\left(\frac{\pi T_{WS}}{\sqrt{2} T }\right)} \nonumber \\
 &\quad\quad\quad\left[1+O\left(\frac{1}{g_M^2},\frac{g_M^3 B}{m_M^2},\frac{g_M^2T}{m_M},\frac{g_M T^2}{B}\right)\right],\label{eq:rate_intro_magnetic}
\end{align}
where $s$ is the spin of the magnetic monopole. The expression applies for temperatures satisfying $T>T_{WS}\approx\left(4 g_M B^3/\pi^3 m_M^2\right)^{1/4}$. Note again that there are now only four dimensionless combinations of parameters in the big $O$, rather than five.

What do our results mean for magnetic monopole searches? If the magnetic field and temperature in a given heavy ion collision vary slowly on the time and length scales of the sphaleron, one may make the approximation that the magnetic field and temperature are locally constant. That is, locally, the rate of pair production is approximated by the rate derived for a constant field and temperature. In this case the cross section for pair production of magnetic monopoles in heavy-ion collisions, $\sigma_{M\bar{M}}$, is given by
\begin{align}
 \frac{\mr{d}\sigma_{M\bar{M}}(\sqrt{s},b)}{\mr{d}b} \approx& \frac{\mr{d}\sigma^{\mr{inel}}_{HI}(\sqrt{s},b)}{\mr{d}b} \nonumber \\ & \int \mathrm{d}^4 x \   \Gamma(\bar{B}(x;\sqrt{s},b),\bar{T}(x;\sqrt{s},b)), \label{eq:crosssection_def}
\end{align}
where $\sqrt{s}$ is the centre of mass energy, $b$ is the impact parameter, $\sigma^{\mr{inel}}_{HI}$ is the total inelastic cross section for the specific heavy-ion collision and $\bar{B}(x;\sqrt{s},b)$ and $\bar{T}(x;\sqrt{s},b)$ are the event-averaged, local magnetic field and temperature at a point $x$ in the fireball for events with centre of mass energy $\sqrt{s}$ and impact parameter $b$. The integrals in Eq. \eqref{eq:crosssection_def} can all be done in the saddle point approximation.


Due to the strong coupling of magnetic monopoles, Eq. \eqref{eq:rate_intro_magnetic} is unfortunately valid only for parametrically low temperatures, $T/m_M\ll 1/g_M^2\sim e^2$, and weak magnetic fields, $B/m_M^2\ll 1/g_M^3\sim e^3$. This is due to the requirement that the self-force instability be moved to much shorter length scales than those of the sphaleron, a hierarchy which breaks down when the rate, and consequently the cross section, are still extremely exponentially suppressed,
\begin{equation}
 \frac{\sigma_{M\bar{M}}}{\sigma^{\mr{inel}}_{HI}} \sim  \ \mr{e}^{- c/e^2},\label{eq:suppression}
\end{equation}
for some $c>0$, with $c=O(1)$. This same factor has been argued to suppress magnetic monopole pair production in high energy particle collisions \cite{witten1979baryons,drukier1982monopole}. However, in our case the arguments of Refs. \cite{witten1979baryons,drukier1982monopole} do not apply and we have no reason to expect the same exponential suppression at higher temperatures. Rather Eq. \eqref{eq:suppression} results due to the limitations of our approximation scheme as applied to strongly coupled magnetic monopoles. Further theoretical work is needed to go beyond this regime.

At higher temperatures than $O(e^2 m_M)$, the inverse temperature becomes smaller than the classical radius of a magnetic monopole, $r_c\sim g_M^2/(4\pi m_M)$, and in considering fluctuations of the sphaleron, the structure of the magnetic monopoles cannot be ignored. However, as the sphaleron is independent of the Euclidean time direction, one might still expect the worldline sphaleron to capture the exponential dependence of the rate.

For 't Hooft-Polyakov monopoles this can be checked explicitly. In this case, the temperature $e^2 m_M$ is a factor of $e$ smaller than the critical temperature, $T_c\sim e m_M$, at which the symmetry of the vacuum is restored. Thus, if one were to perform a semiclassical calculation of the thermal Schwinger rate directly in the field theory, one should be able to calculate the rate up to $T \sim T_c \sim e m_M$, so gaining a factor of $1/e$ on the worldline calculation. In this case one should be able to calculate the rate beyond the semiclassical approximation following the approach of Refs. \cite{Moore:1998swa,Moore:2001vf,Moore:2000jw}. A classical, stochastic equation will govern the real-time dynamics of the sphaleron, and hence determine the prefactor \cite{Arnold:1996dy,Bodeker:1998hm}.

Above $T_c$ in a grand unified theory one can no longer define a magnetic field and magnetic monopoles can no longer be thought of as particles. However, if such a system were to cool, magnetic monopoles would be produced via the freezing out of thermal fluctuations \cite{rajantie2002magnetic}.

\section{Conclusions\label{sec:prefactor_summary}}

The worldline sphaleron describes thermal Schwinger pair production at sufficiently high temperature ($T>T_{WS}$) in Abelian gauge theories, regardless of the strength of the coupling. The rate of this process, Eq. \eqref{eq:rate_intro}, is the chief result of this paper.

In order to work to all orders in $g$, we were forced to restrict our calculation to sufficiently heavy charged particles. In this regime the rate is exponentially suppressed, by the Boltzmann factor, $2m_*/T\approx 2m/T$. For weakly coupled electric particles, we must have $m/T\gg 1$, which is not too restrictive and leads to the exciting possibility of an experimentally observable rate of pair production. Further, in its region of validity, the worldline sphaleron rate turns out to be much faster than both perturbative photon fusion and the zero temperature Schwinger rate. However, for magnetic monopoles, the validity of our approximations require instead that $m/T\gg g^2\geq g_D^2\approx 430$, and hence the results derived in this paper are unfortunately too suppressed to be directly applicable to monopole searches in heavy-ion collisions.

For larger values of $T/m$ or $g^3 B/m^2$, the exponential suppression of the rate, Eq. \eqref{eq:action}, is greatly reduced, leading to predictions of a measurable rate of pair production. It was a regime where $g^3 B/m^2=O(1)$ that led to the mass bounds in Ref. \cite{gould2017magnetic}. However, our calculation of the prefactor breaks down in this case, due to the presence of the self-force instability. One might hope that the calculated exponential suppression still gives a good approximation to the rate at $O(1)$ values of $g^3 B/m^2$, though the corrections to this cannot be calculated within our approach.


\section*{Acknowledgements}
OG would like to thank Sergey Sibiryakov and Toby Wiseman for useful discussions. AR is supported by STFC grant ST/P000762/1 and OG was supported first by an STFC studentship and then by the Research Funds of the University of Helsinki.

\bibliography{sphaleronPrefactor5}

\end{document}